\documentclass[11pt,twocolumn,tight,times]{aastex62}
\usepackage{graphicx,color}
\usepackage{subfigure}
\usepackage{mathrsfs,amsmath}
\usepackage{natbib}
\usepackage{ulem}
\usepackage{hyperref}
\usepackage{CJK}
\emergencystretch=\maxdimen
\begin{document}
\begin{CJK*}{UTF8}{gbsn}

\title{Multiband Gravitational Wave Detection Prospects for M31 UCXB-1 System in Low and Middle Frequency Band}
\shorttitle{Multiband GW for M31 UCXB-1 }
\shortauthors{Guo, et al.}
\author[0000-0001-5174-0760]{Xiao Guo (郭潇)}\thanks{guoxiao@nao.cas.cn}
\affil{Institute for Gravitational Wave Astronomy, Henan Academy of Sciences, Zhengzhou 450046, Henan, China}

%
\author[0000-0002-1932-7295]{Zhoujian Cao (曹周键)}\thanks{zjcao@amt.ac.cn}
\affil{School of physics and astronomy, Beijing Normal University, 19 Xinjiekouwai St, Beijing 100875,China}
\affil{School of Fundamental Physics and Mathematical Sciences, Hangzhou Institute for Advanced Study, University of Chinese Academy of Sciences, No.1 Xiangshan Branch, Hangzhou 310024, China}
\author[0000-0001-7952-7945]{Zhiwei Chen (陈智威)}
\affil{National Astronomical Observatories, Chinese Academy of Sciences, 20A Datun Road, Beijing 100101, China}
\affil{School of Astronomy and Space Science, University of Chinese Academy of Sciences, 19A Yuquan Road, Beijing 100049, China}

%



\begin{abstract}

The recent discovery of M31 UCXB-1, the first extragalactic ultracompact X-ray binary (UCXB) with an orbital period of $T_{\rm orb} \sim 465$ s, presents a unique laboratory for studying close binary evolution and an unprecedented target for continuous gravitational wave (GW) searches. Its identification as a strong candidate black hole-white dwarf (BH-WD) system, combined with its exceptionally short period and high X-ray luminosity, suggests it may be one of the most vital low-frequency GW sources in M31. In this paper, we investigate the detectability of its GW signal for future space-borne detectors in multiband GW detection. We find that while its signal-to-noise ratio (S/N) for low-frequency detectors remains marginal for high-confidence detection, middle-frequency detectors such as DECIGO and BBO are far more promising, potentially achieving S/N $\varrho>8$ within reasonable observational duration. 
With a primary mass of only $m_1 > 5.4M_\odot$ (or $6.6M_\odot$), the network of all low and middle frequency detector (or BBO alone) is capable of detecting GW from this system with a $\varrho > 8$, during 10-year observation.
Furthermore, orbital eccentricity can enhance the GW strain at higher harmonics, further improving its detectability, especially for middle-frequency detectors.
This study establishes M31 UCXB-1 as a key prototype of short-period UCXBs, cementing its role as a cornerstone for multiband, multi-messenger astrophysics and a vital bridge between X-ray astronomy and the future GW era.


\end{abstract}


\section{Introduction}
\label{sec:intro}

Ultracompact X-ray binaries (UCXBs), with orbital periods below $\sim 80$ minutes, represent the end state of low-mass X-ray binary evolution involving a compact accretor [neutron star (NS) or black hole (BH)] and a hydrogen-deficient donor \citep{2021MNRAS.506.4654W,2023pbse.book.....T}. They are powerful laboratories for studying extreme accretion physics, tidal interactions, and angular momentum loss mechanisms \citep[e.g.,][]{2016ApJ...830..131C, 2021MNRAS.503.3540C}. Furthermore, their tight orbits make them prominent sources of persistent, low-frequency gravitational waves (GWs) driven by radiation reaction, placing them squarely within the sensitivity bands of proposed space-borne interferometers \citep{2009CQGra..26i4030N,2020ApJ...900L...8C,2020ApJ...896..129C, 2023ApJ...944...83Q,2021MNRAS.503.5495S, 2023ApJ...953..153H,2025ApJ...995...27X}. including  Laser Interferometer Space Antenna (LISA, \citealt{2017arXiv170200786A}), Tianqin \citep{2016CQGra..33c5010L}, and Taiji \citep{2017SSPMA..47a0404H}. 

Recently, a remarkable system has emerged from deep X-ray monitoring of the Andromeda galaxy (M31). \citet{2024MNRAS.530.2096Z} reported the discovery of a periodic X-ray source (RX J0042.3+4115/3XMM J004222.9+411535) in the M31 bulge with a period of $T_{\rm orb} \approx 465$\,s and a peak luminosity of $L_{X} \sim 2.9 \times 10^{38}$ erg s$^{-1}$. Follow-up spectral analysis by \citet{Ma_Zhang2026} strongly argues that this source, designated M31 UCXB-1, is the first known extragalactic UCXB and a compelling BH-white dwarf (WD) candidate. It possesses the shortest orbital period among known UCXBs. 
The unique parameters of M31 UCXB-1—a likely BH-WD binary with a $\sim 10^2$-s period at a distance of 785 kpc—make its GW emission exceptionally interesting. The expected GW frequency ($f_{\rm gw}\approx 4.3$ mHz) falls within the sensitivity range of low-frequency GW observatories. Previous estimates suggest it may be detectable by LISA/Taiji if the BH mass is sufficiently high \citep{2025ApJ...986..219Y,Ma_Zhang2026}. However, a detailed and systematic assessment of its detectability across the full landscape of future GW missions, considering middle-frequency detectors, mission durations, parameter estimation error and the potential impact of orbital eccentricity, is still lacking.

In this paper, we perform a comprehensive investigation into the GW detectability of M31 UCXB-1 in low and middle frequency band. We go beyond simple characteristic strain comparisons and compute expected signal-to-noise ratios (S/N) for a suite of planned low and middle frequency detectors, including LISA, Taiji, TianQin, DECi-hertz Interferometer Gravitational wave Observatory (DECIGO) \citep{2006CQGra..23S.125K}, and Big Bang Observer (BBO) \citep{2006CQGra..23.4887H}. We explore a realistic range of system parameters (BH mass, orbital eccentricity) informed by the latest observational and theoretical constraints \citep{2024MNRAS.530.2096Z,Ma_Zhang2026, 2025ApJ...986..219Y}. Our analysis reveals that middle-frequency detectors offer a superior pathway for a high-confidence detection of this source. Furthermore, we quantify that if the system exhibits non-zero orbital eccentricity, it can enhance the detection S/N for both low- and middle-frequency GW detectors, with a particularly pronounced improvement for middle-frequency detectors.
This work solidifies M31 UCXB-1's role as a prime multimessenger and multiband target and provides critical predictions for the observing strategies of forthcoming GW observatories in low and middle frequency band.

This paper is structured as follows: In Section~\ref{sec:GWdet}, we assess the GW detectability of M31 UCXB-1 and the associated parameter estimation uncertainties for low- and middle-frequency GW detectors, under the assumption of a circular orbit. Subsequently, in Section~\ref{sec:ecc}, we extend the analysis to include eccentric orbital configurations. Finally, we summarize our findings and present conclusions in Section~\ref{sec:concl}.

\section{GW Detectability Analysis}
\label{sec:GWdet}

\subsection{Parameter Setting}
The binary system is located at a distance of approximately 785 kpc from Earth. Its orbital period is about 465 s, corresponding to an orbital frequency of \(2.15 \times 10^{-3}\) Hz. The WD in the system has a mass of \(0.08761\,M_\odot\), while the mass of the black hole remains uncertain due to its degeneracy with the orbital inclination in X-ray observations \citep{Ma_Zhang2026}.
A BH mass of $m_1 = 3M_\odot$ corresponds to an inclination $\iota \geq 73.5^{\circ}$, while $m_1 = 20M_\odot$ corresponds to $\iota \geq 86.7^{\circ}$ \citep{Ma_Zhang2026}. Since the amplitude of GW is not sensitive to the inclination $\iota$ (see Table~\ref{tab:gw}), and eclipsing phenomena reflects this system is basically edge on. Unless otherwise specified, throughout this work we adopt an edge-on orbital inclination of $\iota = 90^{\circ}$.
Some fixed parameters about the system are summarized in Table~\ref{tab:para}. Since \citet{Ma_Zhang2026} assume that its orbit is circular, we just consider circular orbit in this Section, and we will consider its eccentric orbit in next Section (Sec.~\ref{sec:ecc}). 
\begin{table}
\caption{Some parameter settings for M31 UCXB-1  system.}
\begin{center}
\begin{tabular}{ccccc}
\hline \hline
$d$\,(kpc) & $m_2(M_\odot)$ & $T_{\rm orb}$(s) & $f_{\rm orb}$(Hz) &  \\ \hline
785 & 0.08761 & 465 & $2.15\times10^{-3}$ & \\

 \hline \hline
\end{tabular}
\end{center}
\label{tab:para}
\end{table}

\subsection{GW Signals}
The GW radiated from a binary system can be expressed as \citep{maggiore2008gravitational,2025ApJ...978..104G}
\begin{eqnarray}
h_+ & = & A_+ h_0\cos\Phi(t), 
\label{eq:hplus}\\
h_\times & = & A_\times h_0\sin\Phi(t).
\label{eq:hcross}
\end{eqnarray}
where $h_0=\frac {4{G}^{5/3}\mathcal{M}^{5/3}(\pi f)^{2/3}}{{c}^4d}$, $f=2f_{\rm orb}$ is the frequency of GW,
$A_+=\frac{1+\cos^2\iota}{2}$, $A_\times=\cos\iota$, $+,\times$ represent different polarizations, $\mathcal{M}=(m_1m_2)^{3/5}/(m_1+m_2)^{1/5}$ is the chirp mass of this system and $\Phi(t)$ is the phase of GW. 
The root-mean-square (RMS) average of GW strain on all polarization is 
$$
h_{\rm rms}=\sqrt{\langle h_{+}^2+h_{\times}^2\rangle}=h_0\sqrt{\frac{A_{+}^2+A_{\times}^2}{2}}.
$$
Adopting matched filtering method, the characteristic strain for GW is \citep[see][]{2015CQGra..32a5014M, 2005ApJ...623...23S}
\begin{equation}
h_{\rm c}(f)=h_{\rm rms}\sqrt{N_{\rm eff}}, 
\label{eq:h_char}
\end{equation}
where $N_{\rm eff}=\min\left\{\frac{f^2}{\dot{f}}, T_{\rm obs}f\right\}$\citep{2022ApJ...939...55G,2025ApJ...978..104G},
the variation rate of frequency is
$\dot{f}=\frac{96}{5}\pi^{8/3} \left(\frac{G\mathcal{M}}{c^3} \right)^{5/3}f^{11/3}$.
For this binary system, the relation \(\frac{f^2}{\dot{f}} \gg T_{\mathrm{obs}} f\) holds, implying that the evolution timescale of the GW frequency is much longer than the coherent observing duration. Consequently, the characteristic strain \(h_{\mathrm{c}}\) depends primarily on the observation duration \(T_{\mathrm{obs}}\).

Given \( m_1 \gg m_2 \), $\mathcal{M}\propto m_1^{2/5}$. Using Equation~\eqref{eq:h_char},
the characteristic strain can be reduced to
\begin{equation}
h_{\rm c} \;\approx\; 3.76\times10^{-22}\,
\left(\frac{m_1}{3M_\odot}\right)^{2/3}
\left(\frac{T_{\rm obs}}{10\ \mathrm{yr}}\right)^{1/2}.
\label{eq:hc}
\end{equation}

And its merger time scale due to GW radiation can be estimated by \citep{maggiore2008gravitational}
\begin{equation}
t_{\rm gw}\approx2.32\times10^{5}{\rm yr}\left(\frac{m_1}{3M_\odot}\right)^{-2/3}.
\label{eq:t_m}
\end{equation}

Therefore, we can summarize several typical characteristic strains and other dynamical parameters for different cases in Table~\ref{tab:gw}. Although the accretion process of a black hole can alter the masses $m_1$ and $m_2$ of the binary system, the resulting mass change is negligible over the observational timescale ($<10$\,yr) due to low accretion rate [$\dot{M}\sim10^{-8}M_\odot$/yr \citep{2025ApJ...986..219Y}]. Therefore, in our study on GWs, we neglect the mass variation due to accretion as well as the potential changes in orbital dynamics.
\begin{table*}
\caption{GW emission properties and orbital dynamics for 3 binary system configurations, showing the characteristic strain amplitudes for observation periods of 4 yr (\(h_c\)) and 10 yr (\(h_e\)), the GW frequency derivative (\(\dot{f}_{\mathrm{gw}}\)), and the merger timescale (\(t_{\mathrm{gw}}\)).}
\begin{center}
\begin{tabular}{ccccccc}
\hline \hline
Cases & $m_1(M_\odot)$ & $\iota(^{\circ})$ & $h_c$( $T_{\rm obs}=4$\,yr) &$h_c$( $T_{\rm obs}=10$\,yr) & $\dot{f}_{\rm gw}(\rm Hz\cdot yr^{-1})$ & $t_{\rm gw}$(yr)\\ \hline
Small inclination & 3 & 73.5 & $4.20\times10^{-22}$ & $6.65\times10^{-22}$ & $6.943\times10^{-9}$ & $2.32\times10^{5}$\\
Low mass & 3 & 90 & $3.76\times10^{-22}$ &  $5.95\times10^{-22}$& $6.943\times10^{-9}$ & $2.32\times10^{5}$\\

Massive & 20 & 90 & $1.34\times10^{-21}$ &  $2.12\times10^{-21}$ & $2.48\times10^{-8}$ & $6.48\times10^{4}$\\
 \hline \hline
\end{tabular}
\end{center}
\label{tab:gw}
\end{table*}

\subsection{GW Detectors}

\begin{table}
\caption{Parameter settings for different GW detectors.}
\begin{center}
\begin{tabular}{cccc}
\hline \hline
GW detector& $f_{\min}$(Hz) & $f_{\max}$(Hz) & References \\ \hline
Tianqin & $10^{-4}$ & 0.5 &\cite{2020PhRvD.101j3027L} \\
LISA & $10^{-4}$ & 0.5 &\cite{2019CQGra..36j5011R}\\
Taiji & $10^{-4}$ & 0.5 &\cite{2023PhRvD.107f4021L} \\
DECIGO & $10^{-3}$ & $10^2$ &\cite{2011PhRvD..83d4011Y,2017PhRvD..95j9901Y} \\
BBO & $10^{-3}$ & $10^2$ &\cite{2011PhRvD..83d4011Y,2017PhRvD..95j9901Y}  \\

 \hline \hline
\end{tabular}
\end{center}
\label{tab:detector}
\end{table}

\begin{figure}
\centering
\includegraphics[width=0.48\textwidth]{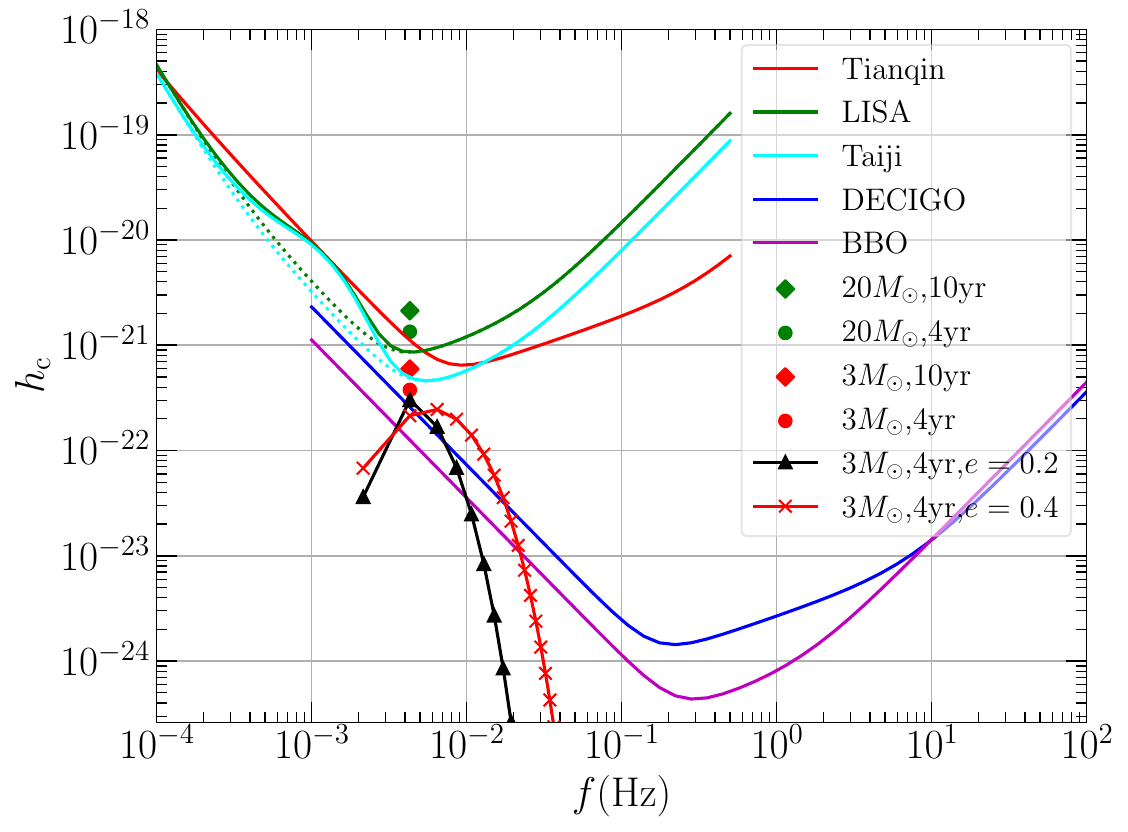}

\caption{Possible characteristic strain $h_{\rm c}(f)$ of GW emitted from this system [red ($m_1=3M_\odot$) and green ($m_1=20M_\odot$) points($T_{\rm obs}=4$\,yr)/diamond($T_{\rm obs}=10$\,yr)] and noise curve $h_{\rm n}(f)$ for different GW detectors in low and middle frequency band with different colors represent different GW detectors as shown in the legend. These dotted lines for Taiji and LISA represent the noise curves of GW detectors without Galactic binaries foreground confusion \citep[See][]{2023PhRvD.107f4021L}. And this black line with triangle (red line with $\times$) symbols represents characteristic strain $h_{{\rm c},i}$ for eccentric orbit with $e=0.2$(0.4), $m_1=3M_\odot$, and $T_{\rm obs}=4$\,yr.
}
\label{fig:SenCur}
\end{figure}

It is likely that this system will be detected by future low-frequency and middle-frequency GW detectors. The low-frequency GW detectors encompass LISA, Taiji, and Tianqin, whereas the middle-frequency GW detectors include DECIGO and BBO in this paper. For all GW detectors used in this work, their noise power spectral density $S_{\rm n}(f)$ are summarized in the references provided in Table~\ref{tab:detector} and their detectable frequency ranges are also summarized in Table~\ref{tab:detector}. We show some typical characteristic strain values of this GW source (red and green points/diamonds, Eq.~\ref{eq:hc}) and noise curve of all kinds of GW detectors in low and middle frequency band in the Figure~\ref{fig:SenCur}. 
The characteristic strain values computed by us are broadly consistent with those reported by \citet{2025ApJ...986..219Y,Ma_Zhang2026}. The slightly lower amplitude in our results is due to our choice of orbital inclination $\iota = 90^{\circ}$.

It is worth noting that although the confusion noise from Galactic foreground binaries in the noise curve is calculated for an observation time $T_{\rm obs}=4$\,yr, their frequencies are lower than that of M31 UCXB-1 and therefore do not affect its detection, especially when $T_{\rm obs}>4$\,yr. Thus we still adopt the same noise curves for LISA and Taiji even if $T_{\rm obs}\neq4$\,yr.

\subsection{S/N}
\begin{figure}
\centering
\includegraphics[width=0.48\textwidth]{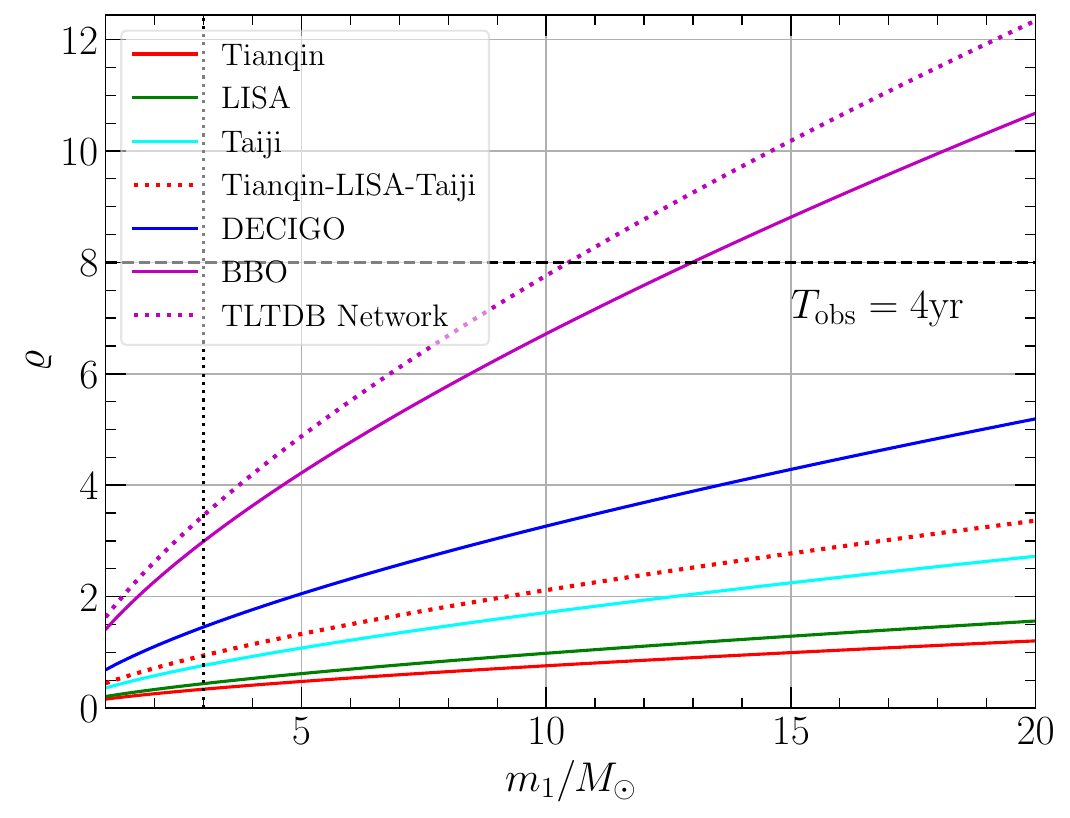}

\caption{S/N as the function of primary mass $m_1$ for $T_{\rm obs}=4$\,yr, where the vertical black dotted line indicates the lower mass limit for black holes, $3M_\odot$; the horizontal black dashed line indicates threshold S/N $\varrho=8$. We utilize colorful dotted lines to represent S/N for GW detector networks.
}
\label{fig:SNR_m4}
\end{figure}

\begin{figure}
\centering
\includegraphics[width=0.48\textwidth]{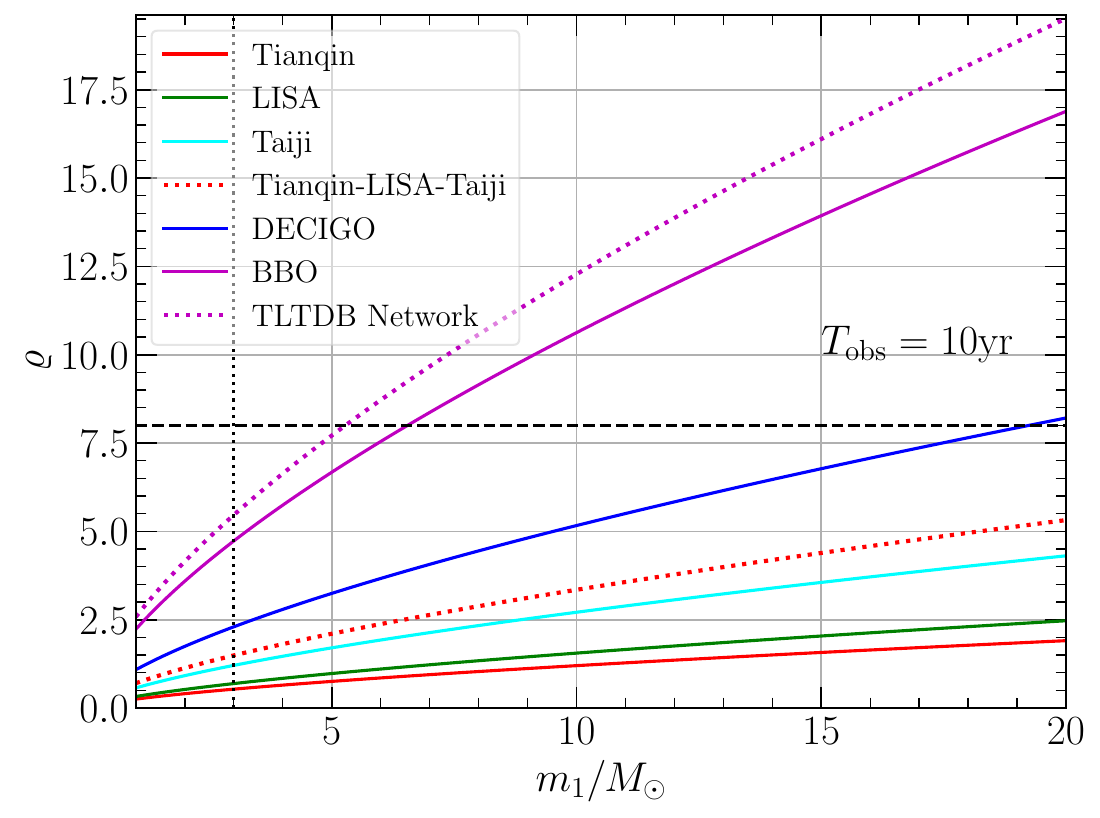}

\caption{S/N as the function of primary mass $m_1$ for $T_{\rm obs}=10$\,yr. It is similar to Figure~\ref{fig:SNR_m4} except $T_{\rm obs}=10$\,yr.
}
\label{fig:SNR_m10}
\end{figure}

The S/N of GW from this system detected by a GW detector is \citep{2015CQGra..32a5014M, 2015CQGra..32e5004M}
\begin{equation}
\varrho=\frac{h_{\rm c}(f)}{h_{\rm n}(f)},
\label{eq:SNR}
\end{equation}
where $h_{\rm n}(f)=\sqrt{fS_{\rm n}(f)}$ represents the noise curve of GW detector, which is shown in Figure~\ref{fig:SenCur}. 
By combining its characteristic strain (Eq.\ref{eq:hc}) with the detector noise, we can compute its S/Ns for different GW detectors. 
It is straightforward to obtain the following relation
\begin{equation}
\varrho\propto m_1^{2/3}T_{\rm obs}^{1/2}.
\label{eq:rho_mT}
\end{equation}
For a network of GW detectors, its effective S/N is calculated as follows\citep{2022ApJ...939...55G,2025ApJ...978..104G}
\begin{equation}
\varrho=\sqrt{\sum_{i=1}^{N_{\rm d}}\varrho_i^2},
\label{eq:network}
\end{equation}
where $\varrho_i$ is the S/N for each detector $i$, the sum is taken over the \(N_{\rm d}\) detectors.
In this paper, we mainly consider two GW detector networks: Tianqin-LISA-Taiji low frequency GW detector network and Tianqin-LISA-Taiji-DECIGO-BBO (TLTDB) network including all low and middle frequency GW detectors in this paper.
Through calculation, we obtain its S/Ns for different detectors as a function of $m_1$, as shown in Figure~\ref{fig:SNR_m4} ($T_{\rm obs}=4$\,yr) and Figure~\ref{fig:SNR_m10} ($T_{\rm obs}=10$\,yr) [See also Figure~\ref{fig:SNR_T3} ($m_1=3M_\odot$) and Figure~\ref{fig:SNR_T20} ($m_1=20M_\odot$) in Appendix~\ref{sec:rho_T}, which demonstrates S/Ns as a function of observation time $T_{\rm obs}$ for different $m_1$ and detectors.]. If the primary object is an astrophysical black hole, its mass should be greater than $3M_\odot$, placing it in the region to the right of the black vertical dotted line in Figure~\ref{fig:SNR_m4} and \ref{fig:SNR_m10}.

From these figures, it can be observed that even under the optimistic conditions of $T_{\rm obs}=10$\,yr and $m_1=20M_\odot$, the S/N for detecting it with the Tianqin-LISA-Taiji network (red dotted lines in Figure~\ref{fig:SNR_m4}, \ref{fig:SNR_m10}, \ref{fig:SNR_T3}, \ref{fig:SNR_T20}) remains less than 8. While for the powerful TLTDB detector network, as long as $m_1>5.4(10.6) M_\odot$, it can be detected with an S/N of 8 over a 10(4)-year observation period.

%
%
%

\subsection{Uncertainty of Parameters}
\label{sec:para}

In GW data analysis, we usually utilize Fisher information matrix to estimation the error or uncetrainty of parameters \citep{2020MNRAS.496..182L,2023MNRAS.522.2951Z,2025arXiv251213151Z}.
For two parameters $p_\alpha$, $p_\beta$ of GW system, the Fisher information matrix (FIM) is defined as follows:
\begin{equation}
	\Gamma_{\alpha\beta} = \left(\frac{\partial h}{\partial p_\alpha} \left|\frac{\partial h}{\partial p_\beta}\right.\right)
	= 2\int \frac{\frac{\partial \tilde{h}}{\partial p_\alpha}\frac{\partial \tilde{h}^*}{\partial p_\beta}+\frac{\partial \tilde{h}}{\partial p_\beta}\frac{\partial \tilde{h}^*}{\partial p_\alpha} }{S_{\rm n}(f)} \, df,
	\label{eq:FIM}
\end{equation}
where $\tilde{h}(f)$ is GW waveform in frequency domain. 

The covariance matrix used to estimate parameter errors is defined as follows:
$$
\langle\delta p_\alpha\delta p_\beta\rangle=(\Gamma^{-1})^{\alpha\beta},
$$
where $\Gamma^{-1}$ is the inverse of FIM. Thus the uncertainties of parameters $p_\alpha$ can be esitmated by 
$$
\Delta p_\alpha=\sqrt{(\Gamma^{-1})^{\alpha\alpha}}.
$$

Many parameters (including $m_2$, $f_{\rm orb}$, distance $d$ and so on) of M31 UCXB-1 system have already been well constrained by X-ray observations, with the primary remaining uncertainties lying in the $m_2$ and the orbital inclination $\iota$. The inclination is constrained into a very narrow parameter space (the system is essentially edge-on $\iota\rightarrow90^{\circ}$) and its effect on the GW amplitude is relatively small. Therefore, the dominant source of uncertainty in our analysis arises from a free parameter $m_1$, and we ignore the error of other parameters.
Since the amplitude of GW is directly influenced by the chirp mass, we can evaluate the uncertainty in primary mass $\Delta m_1$ by calculating the uncertainty in the chirp mass $\Delta \mathcal{M}$.

For its Fisher information matrix, we consider only the chirp mass as the parameter of interest, thus 
$$
(\Gamma^{-1})_{\mathcal{M}\mathcal{M}}=\frac{1}{\Gamma_{\mathcal{M}\mathcal{M}}},
$$
hence the uncertainty of chirp mass $\mathcal{M}$ is
\begin{equation}
\Delta \mathcal{M}=\frac{1}{\sqrt{ \left(\frac{\partial h}{\partial \mathcal{M}} \left|\frac{\partial h}{\partial \mathcal{M}}\right.\right)}}
=\frac{1}{\sqrt{4\int\frac{|\frac{\partial \tilde{h}}{\partial \mathcal{M}}|^2}{S_{\rm n}(f)}df}},
\label{eq:Delta_M}
\end{equation}
where $|\tilde{h}(f)| \propto\mathcal{M}^{5/6}$ is the waveform of its GW during its inspiral stage in frequency domain  \citep{1996PhRvD..53.2878F,2018MNRAS.476.2220L}.
Thus it is evident that
$$
\left|\frac{\partial \tilde{h}}{\partial \mathcal{M}}\right|=\frac{5}{6}\frac{|\tilde{h}|}{\mathcal{M}}.
$$
Substitute it into Equation~\eqref{eq:Delta_M}, finally we obtain
\begin{equation}
\Delta \mathcal{M}=\frac{6\mathcal{M}}{5\sqrt{4\int\frac{|\tilde{h}|^2}{S_{\rm n}(f)}df}}=\frac{6\mathcal{M}}{5\varrho},
\end{equation}
where S/N $\varrho=\sqrt{4\int\frac{|\tilde{h}|^2}{S_{\rm n}(f)}df}$ is another equivalent definition of S/N.
Furthermore, since $m_1\gg m_2$, $\mathcal{M}\approx m_1^{2/5}m_2^{3/5}$, the uncertainty of primary mass $\Delta m_1$ is
\begin{equation}
\Delta m_1=\Delta \mathcal{M}\left(\frac{\partial \mathcal{M}}{\partial m_1}\right)^{-1}\approx\frac{5}{2}\left(\frac{m_1}{m_2}\right)^{3/5}\Delta \mathcal{M}=\frac{3m_1}{\varrho}.
\end{equation}

\begin{figure}
\centering
\includegraphics[width=0.48\textwidth]{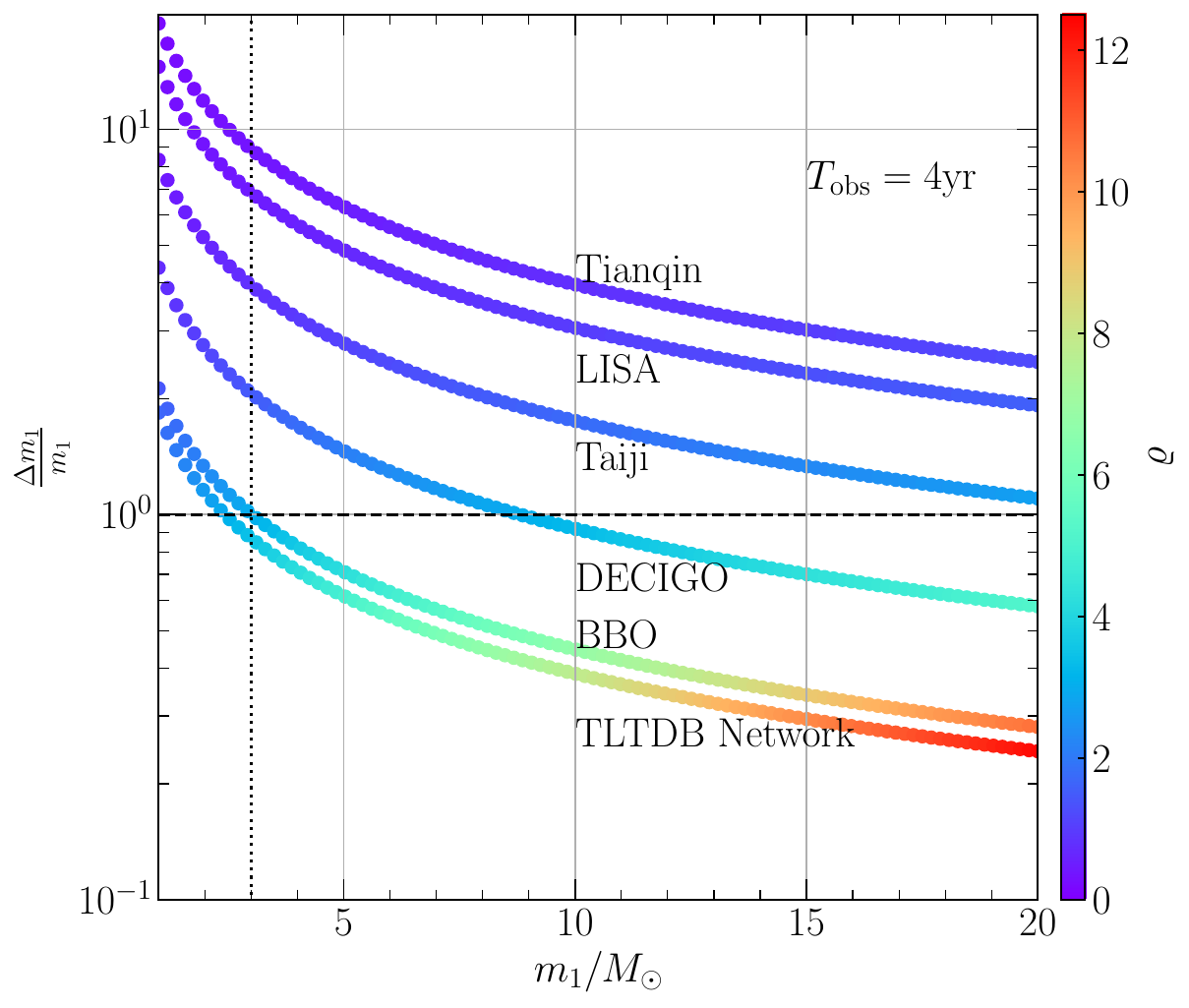}

\caption{The relative uncertainty $\Delta m_1/m_1$ versus BH mass $m_1$ for $T_{\rm obs}=4$\,yr. We utilize different colors to represent different values of the S/N as indicated in color bar. Since a larger \(m_1\) leads to a higher S/N $\varrho$, the constraining power on \(m_1\) correspondingly increases, and the relative uncertainty \(\Delta m_1/m_1\) becomes smaller.
}
\label{fig:Dm_m4}
\end{figure}

\begin{figure}
\centering
\includegraphics[width=0.48\textwidth]{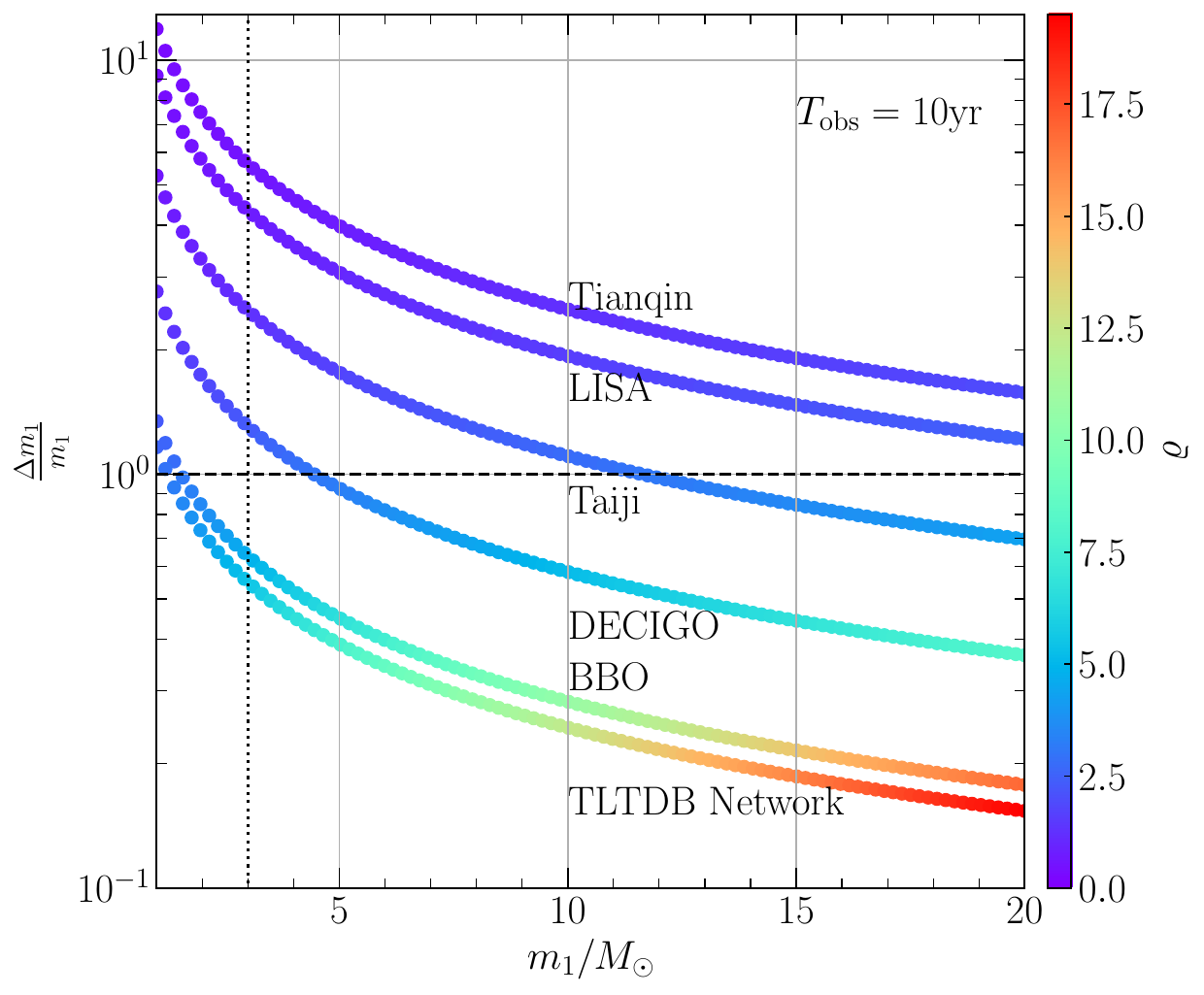}

\caption{The relative uncertainty   $\Delta m_1/m_1$ versus  BH mass $m_1$ for $T_{\rm obs}=10$\,yr. It is similar to Figure~\ref{fig:Dm_m10}, except $T_{\rm obs}=10$\,yr.
}
\label{fig:Dm_m10}
\end{figure}

Therefore, the relative uncertainty of primary mass $m_1$ is
\begin{equation}
\frac{\Delta m_1}{m_1}=\frac{3}{\varrho}.
\label{eq:rDelta_m}
\end{equation}
According to Equation~\eqref{eq:rDelta_m} and \eqref{eq:SNR}, the relative uncertainties about chirp mass $\Delta m_1/m_1$ can be work out, as shown in Figure~\ref{fig:Dm_m4} ($T_{\rm obs}=4$\,yr) and  Figure~\ref{fig:Dm_m10} ($T_{\rm obs}=10$\,yr).
Given $m_1=20M_\odot$, BBO can constrain the relative uncertainty of the \( \Delta m_1/m_1 \) parameter to $\sim30\%(20\%)$ during $T_{\rm obs}=4(10)$\,yr. The TLTDB network exhibits slightly stronger constraining power on the parameter uncertainty compared to BBO.
Although accretion onto the BH can influence the estimation of its mass, the parameter uncertainties derived from our GW predictions are substantially larger than the mass variation induced by BH accretion. Therefore, we also neglect the mass variation due to accretion in this analysis.

\section{Eccentric Orbit}
\label{sec:ecc}

\citet{2025arXiv251213151Z,2025ApJ...995...27X} demonstrate that binary systems residing in globular clusters or nuclear clusters can retain nonzero orbital eccentricity. 

If the orbit of a binary system is not circular but eccentric, the frequencies of the GWs it radiates will no longer be simply twice the orbital frequency $f_{\rm orb}$. Instead, there will exist many higher-order harmonic frequencies at integer multiples of the orbital frequency $nf_{\rm orb}$, $n=1,2,3,...$\citep{1963PhRv..131..435P,maggiore2008gravitational}.  Such a GW signal is easier to analyze in the frequency domain. 

\subsection{GW Signal}
In the frequency domain, the GW power radiated from an eccentric binary system can be expressed as a summation of all harmonics
\begin{equation}
P=\sum_{n=1}^{\infty} P(n),
\end{equation}
where $P(n)$ ($n=1, 2, 3, \cdots$) is the angle-averaged GW power radiated in the $n$th harmonic from the eccentric binary given by \citep{1963PhRv..131..435P,2015PhRvD..92f3010H, maggiore2008gravitational}
\begin{equation}
P(n)=\frac{32G^4 m_{1}^2m_{2}^2 M}{5 c^5 a^5}g(n,e),
\end{equation}
where $M=m_1+m_2$ is the total mass, $a$ is the semimajor axis of orbit (see the lower panel in Figure~\ref{fig:ta_m}),
\begin{equation}
\begin{split}
g(n,e)=&\frac{n^4}{32}\bigg[\Big\{J_{n-2}(ne)-2eJ_{n-1}(ne)\\
&+\frac{2}{n}J_n(ne)+2eJ_{n+1}(ne)-J_{n+2}(ne)\Big\}^2\\
&+(1-e^2)\Big\{J_{n-2}(ne)-2J_n(ne)+J_{n+2}(ne)\Big\}^2\\
&+\frac{4}{3n^2}J^2_n(ne)\bigg],
\end{split}
\end{equation}
and $J_n$ is the $n$th order Bessel function. And the RMS amplitude of the angle-averaged GW strain in the $n$th harmonic is \citep{2015PhRvD..92f3010H}
\begin{equation}
H_{n}=2\sqrt{\frac{32}{5}g(n,e)}\frac{G^{5/3}\mathcal{M}^{5/3}}{c^4d n}(2\pi f_{\rm orb})^{2/3},
\end{equation}
and the angle-averaged GW strain means the RMS of the GW strain averaged over the phase angles and the inclination $\iota$. 
In the special case of a circular orbit, only the 2nd harmonic component $H_{2}$ does not vanish ($H_n|_{n\neq 2}=0$), and the GW radiation is almost monochromatic with a frequency $f=2 f_{\rm orb}$. 

And the characteristic strain contributed by the $i$th component for GW from eccentric orbit is $h_{{\rm c},i}\equiv H_{i}\sqrt{f_iT_{\rm obs}}$, and we show two examples with $e=0.2$ (black line with triangles) and $e=0.4$ (red line with $\times$) with $m_1=3M_\odot$ in Figure~\ref{fig:SenCur}.

Since the inclination $\iota$ of this system is unknown, and its value will not significantly influence the S/N of this system, we just consider an angle-averaged GW strain in this Section.

\subsection{S/N}
\label{subsec:SNR}
If adopting the matched-filtering method, the S/N $\varrho$ of this GW signal from eccentric binary can be roughly estimated as \citep{2025ApJ...978..104G}
\begin{equation}
\varrho = \sqrt{ \sum_{i=n_{\rm min}}^{n_{\rm max}}  \frac{ H_{i}^2T_{\rm obs}}{S_{\rm n}(f_i)}} \nonumber =  \sqrt{\sum_{i=n_{\rm min}}^{n_{\rm max}}\frac{h_{{\rm c},i}^2}{h_{\rm n}^2(f_i)}}, 
\label{eq:SNR_e}
\end{equation}
where noise $h_{\rm n}^2(f_i)=f_iS_{\rm n}(f_i)$, $n_{\rm min}=\left[\frac{f_{\rm min}}{f_{\rm orb}}\right]+1$, $n_{\rm max}= \left[\frac{f_{\rm max}}{f_{\rm orb}}\right]$, and $f_{\min}$, $f_{\max}$ for different GW detectors are summarized in the Table~\ref{tab:detector}, and $[x]$ represents the maximum integer not larger than $x$. 
For simplicity, we just assume observation time $T_{\rm obs}=4$\,yr in this Section.

\begin{figure}
\centering
\includegraphics[width=0.48\textwidth]{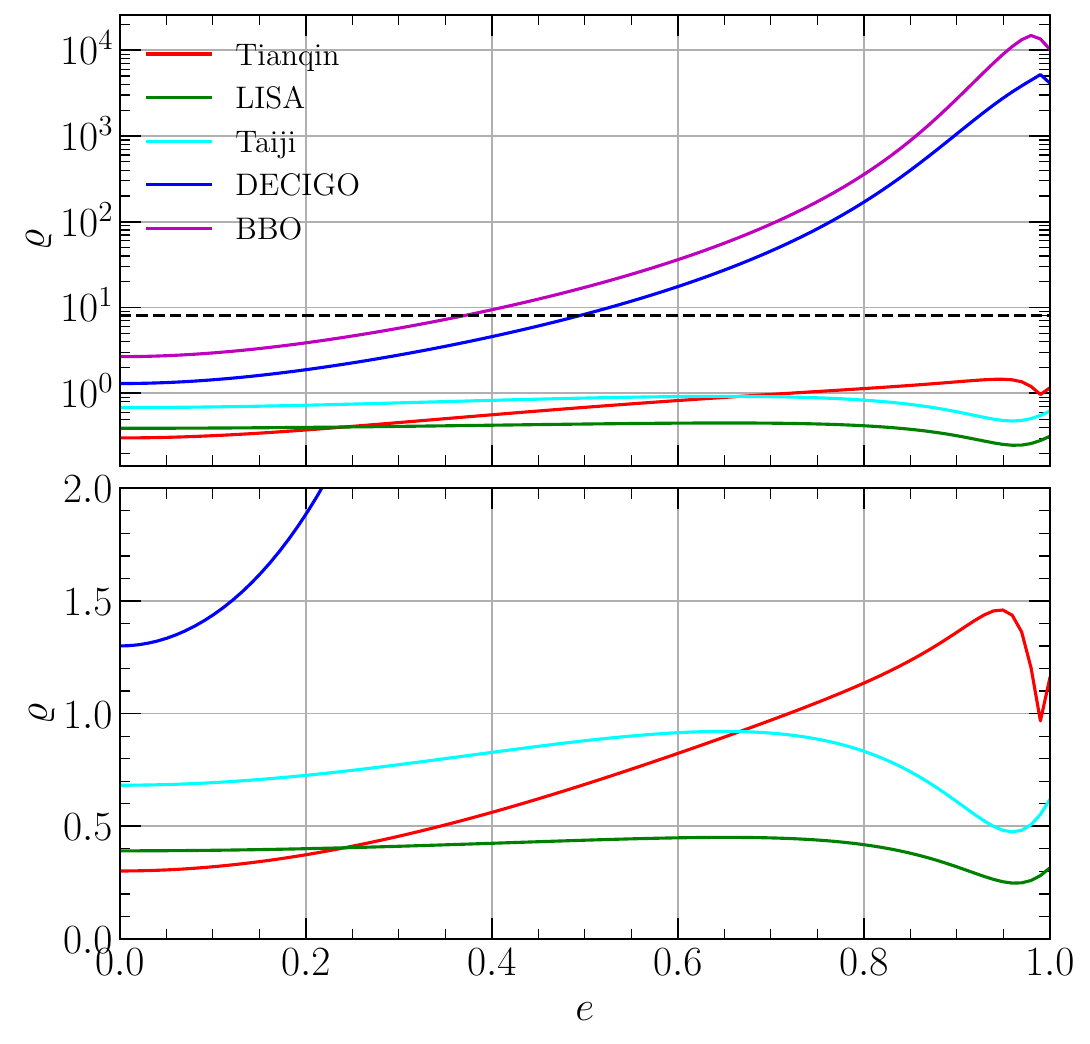}

\caption{S/N as the function of $e$ for $m_1=3M_\odot$, where the horizontal black dashed line indicates threshold S/N $\varrho=8$.
}
\label{fig:SNR_e3}
\end{figure}

\begin{figure}
\centering
\includegraphics[width=0.48\textwidth]{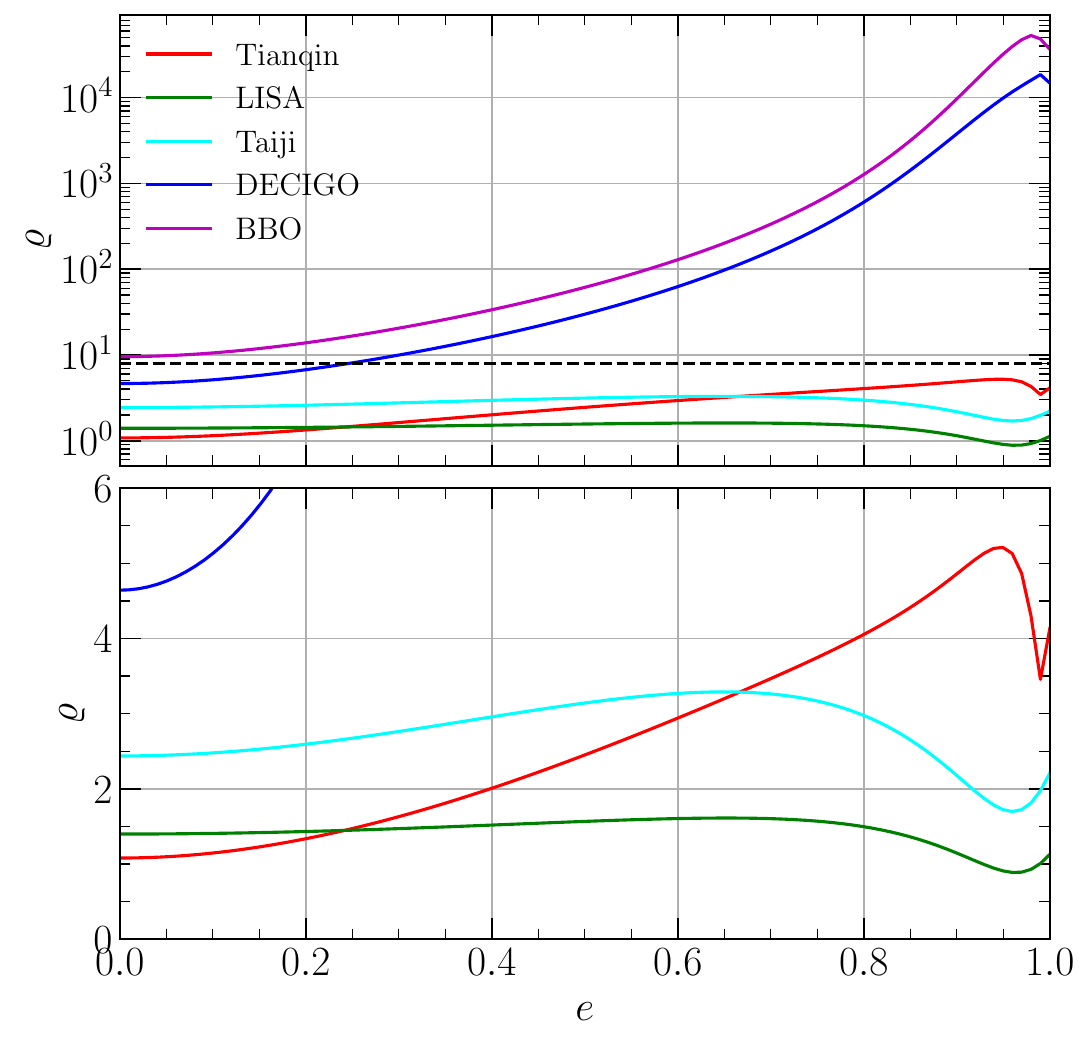}

\caption{S/N as the function of $e$ for $m_1=20M_\odot$.
}
\label{fig:SNR_e20}
\end{figure}

Finally, we can obtain the S/N as the function of eccentricity $e$ in the Figure~\ref{fig:SNR_e3} with $m_1=3M_\odot$ and Figure~\ref{fig:SNR_e20} with $m_1=20M_\odot$. 
It is evident from these figures (\ref{fig:SNR_e3} and \ref{fig:SNR_e20}) that when \(e \rightarrow 0\), the computed S/N values closely match those for a circular orbit (Figure~\ref{fig:SNR_m4}, \ref{fig:SNR_T3}, \ref{fig:SNR_T20}).
These figures show that if the eccentricity of this binary is nonzero, it will be more probable to detect GW from this system with a higher S/N. For almost all these space-based GW detectors, $\varrho$ increases with increasing $e$ as long as $e\lesssim0.7$. For low frequency GW detectors, the increase of S/N is not significant, however, the increase of S/N for middle frequency GW detectors with increasing $e$ is very obvious. 

It should be noted that, we have ignored the evolution of $a$ and $e$, since merger time $t_{\rm gw}\gg T_{\rm obs}$ given $e\lesssim0.9$ (see Figure~\ref{fig:t_e} in Appendix~\ref{sec:t_gw}).
Our estimate of the S/N may become unreliable as \( e \rightarrow 1 \). In this high-eccentricity regime, gravitational-wave emission is significantly enhanced, leading to rapid orbital frequency evolution and a short merger timescale (see Figure~\ref{fig:t_e}), which our simplified steady-signal model does not capture.
However, our calculations demonstrate that if this system has a moderate eccentricity (e.g. $e\gtrsim0.4$ for $m_1=3M_\odot$ or $e\gtrsim0.2$ for $m_1=20M_\odot$), it is probable that it could be detected by middle frequency GW detectors with a very significant S/N and a high confidence.
 
\section{Conclusions}
\label{sec:concl}

Our calculation results demonstrate that although M31 UCXB-1 could be detected by low frequency GW detectors including Taiji, LISA and Tianqin (even the detector network formed by combining all of them) with S/N $\varrho>1$ when primary mass $m_1>10M_\odot$, their S/Ns could not be enough high to claim that this system can be detected with high confidence even if $m_1=20M_\odot$. 
Even if we increase observation time to $T_{\rm obs}=10$\,yr, their S/Ns are still not enough high to claim a high confident detection as long as $m_1\leq20M_\odot$. It should be noted that when the characteristic strain is greater than noise strain $h_c>h_{\rm n}$, it just represents S/N $\varrho>1$, but it can not represent that this GW source can be detected with a high confidence.

However, if we utilize the middle frequency GW detectors such as DECIGO and BBO to detect this potential GW sources, it is possible to detect it with a enough high S/N $\varrho>8$ especially for BBO. 
As long as primary mass $m_1>6.6M_\odot(13M_\odot)\,$, BBO can detect GW from this system with $\varrho>8$ during observation time $T_{\rm obs}=10$\,yr (4\,yr).
If primary mass $m_1>19.4M_\odot$, DECIGO can also detect GW from this system with $\varrho>8$ with observation time $T_{\rm obs}=$10\,yr.
Therefore, middle frequency GW detectors (DECIGO and BBO) are better suited for detecting GWs from such systems with a very short period ($\sim 10^2$\,s) than middle frequency GW detectors.

GW, serving as a complementary probe to electromagnetic observations, provide information inaccessible to electromagnetic observation method alone, such as constraints on the primary mass $m_1$ for M31 UCXB-1 system. As a representative and vital multi-messenger source, M31 UCXB-1 can be observed not only through both GW and electromagnetic channels, but also via GW across different frequency bands (low and middle frequency band), thereby leveraging the strengths of detectors operating in distinct parts of the GW spectrum. Given $m_1=20M_\odot$, BBO can constrain the relative uncertainty of the \( \Delta m_1/m_1 \) parameter to $\sim30\%(20\%)$ during $T_{\rm obs}=4(10)$\,yr.
Multiband joint observations of GW sources can help enhance the S/N of the GW signal, improve localization accuracy, and reduce parameter estimation errors \citep{2020MNRAS.496..182L,2023MNRAS.522.2951Z,2025arXiv251213151Z}. Multiband joint observations in low and middle frequency band represent a promising approach worthy of consideration in future studies of M31 UCXB-1 or similar short-period UCXB with $T_{\rm orb}<10^3$\,s.
In this work, we consider a joint network of all low- and middle-frequency GW detectors, referred to as the TLTDB network.
During a 10(4)-year observation period, the powerful TLTDB detector network can achieve a detection S/N of 8 as long as the GW source with $m_1 > 5.4(10.6)\,M_\odot$.

What is more, we find that if the orbit of this system is not circular but eccentric, it would be more favorable for middle-frequency GW detectors to detect the GWs generated by this system with a higher S/N. 
For a primary mass of \(m_1 = 3M_\odot\), the system becomes detectable with S/N \(\varrho > 8\) by both BBO and DECIGO once the orbital eccentricity satisfies \(e \gtrsim 0.4\). For a higher mass \(m_1 = 20M_\odot\), BBO can achieve \(\varrho > 8\) even for a circular orbit, while DECIGO requires only a modest eccentricity \(e \gtrsim 0.2\) to reach the same detection threshold.

In summary, for such a short-period UCXB (orbital period \(T_{\rm orb} \leq 10^3\) s), detection strategies can extend beyond low-frequency GW observatories. Multiband joint observations employing both low-frequency and middle-frequency detectors offer a promising avenue, potentially enhancing detection confidence, improving parameter estimation, and increasing the overall S/N.

\acknowledgements{
We thank the helpful discussion with Jiachang Zhang, Lijing Shao, Wen-Hong Ruan and Yuetong Zhao.
This work is supported by the National Natural Science Foundation of China under Grant No.~12503001 (X.G.), the National Key Research and Development Program of China under Grant Nos. 2021YFC2203001 (Zhoujian Cao), the National Natural Science Foundation of China under No.~12475049 (Zhoujian Cao) and the Postdoctoral Fellowship Program of CPSF under Grant Number GZB20250735 (Zhiwei Chen).
}
\appendix
\section{S/N as the function of observation time $T_{\rm obs}$}
\label{sec:rho_T}
\begin{figure}
\centering
\includegraphics[width=0.48\textwidth]{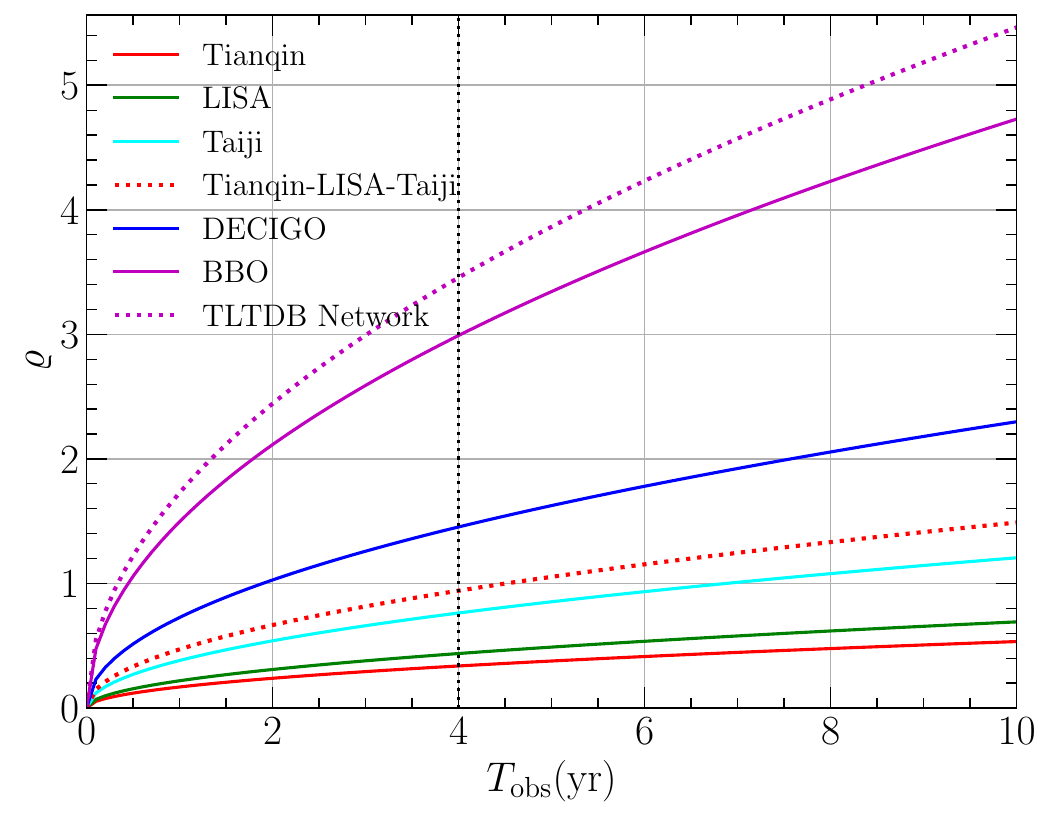}

\caption{S/N as the function of observation time $T_{\rm obs}$ for $m_1=3M_\odot$, where the vertical black dotted line indicates $T_{\rm obs}=4$\,yr.
}
\label{fig:SNR_T3}
\end{figure}

\begin{figure}
\centering
\includegraphics[width=0.48\textwidth]{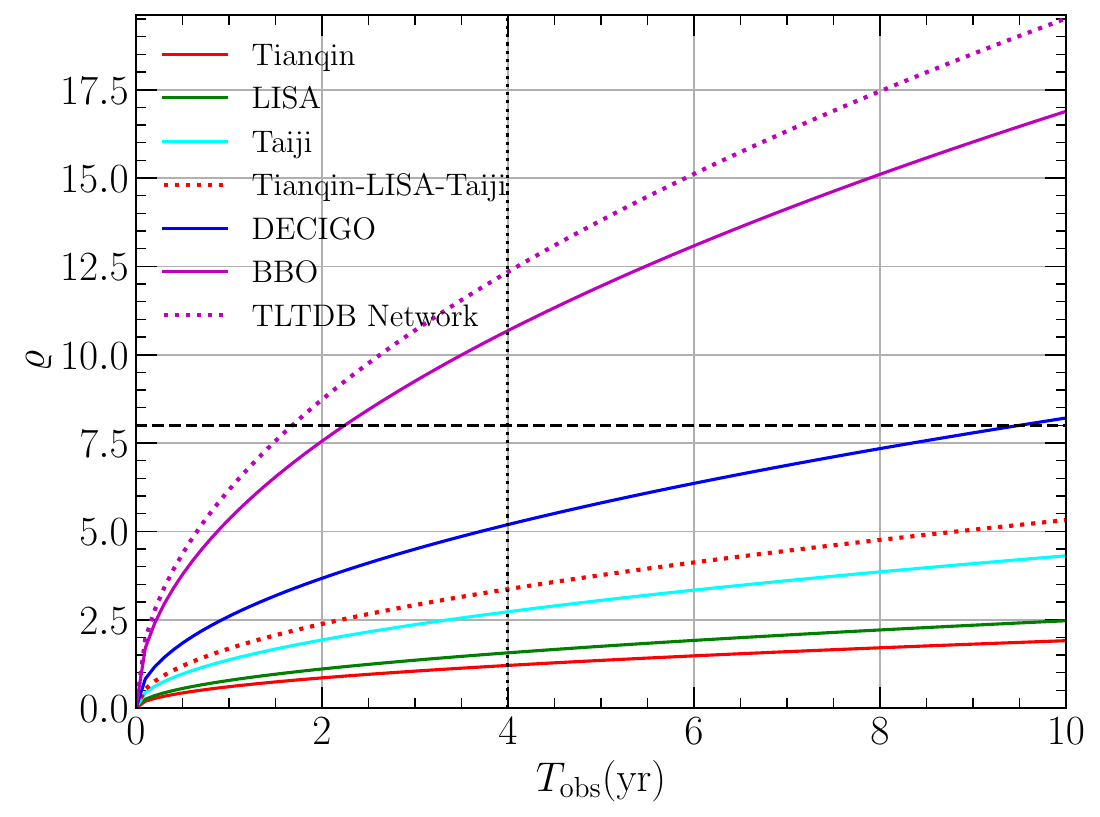}

\caption{S/N as the function of observation time $T_{\rm obs}$ for $m_1=20M_\odot$.
}
\label{fig:SNR_T20}
\end{figure}

Increasing the observation time enhances the S/N for a given GW source, with the relationship described by Equation~\eqref{eq:rho_mT}. We also present its S/N as a function of observation time $T_{\rm obs}$ in Figure~\ref{fig:SNR_T3} ($m_1=3M_\odot$) and Figure~\ref{fig:SNR_T20} ($m_1=20M_\odot$).

\section{Merger time \& semimajor axis}
\label{sec:t_gw}
\begin{figure}
\centering
\includegraphics[width=0.48\textwidth]{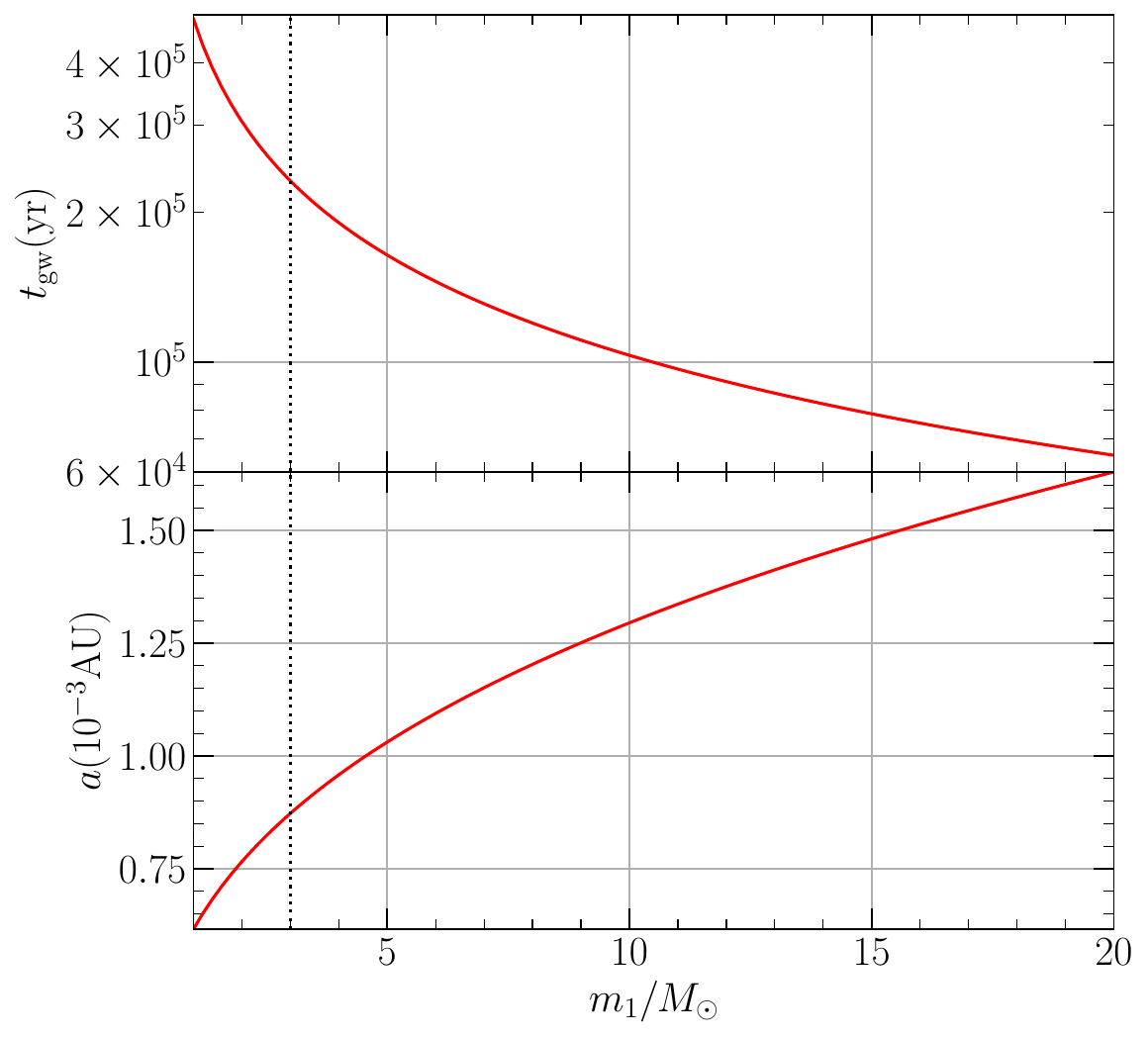}

\caption{The merger time $t_{\rm gw}$ (upper panel) and semi-major axis $a$ (lower panel) as the function of $m_1$.
}
\label{fig:ta_m}
\end{figure}

\begin{figure}
\centering
\includegraphics[width=0.48\textwidth]{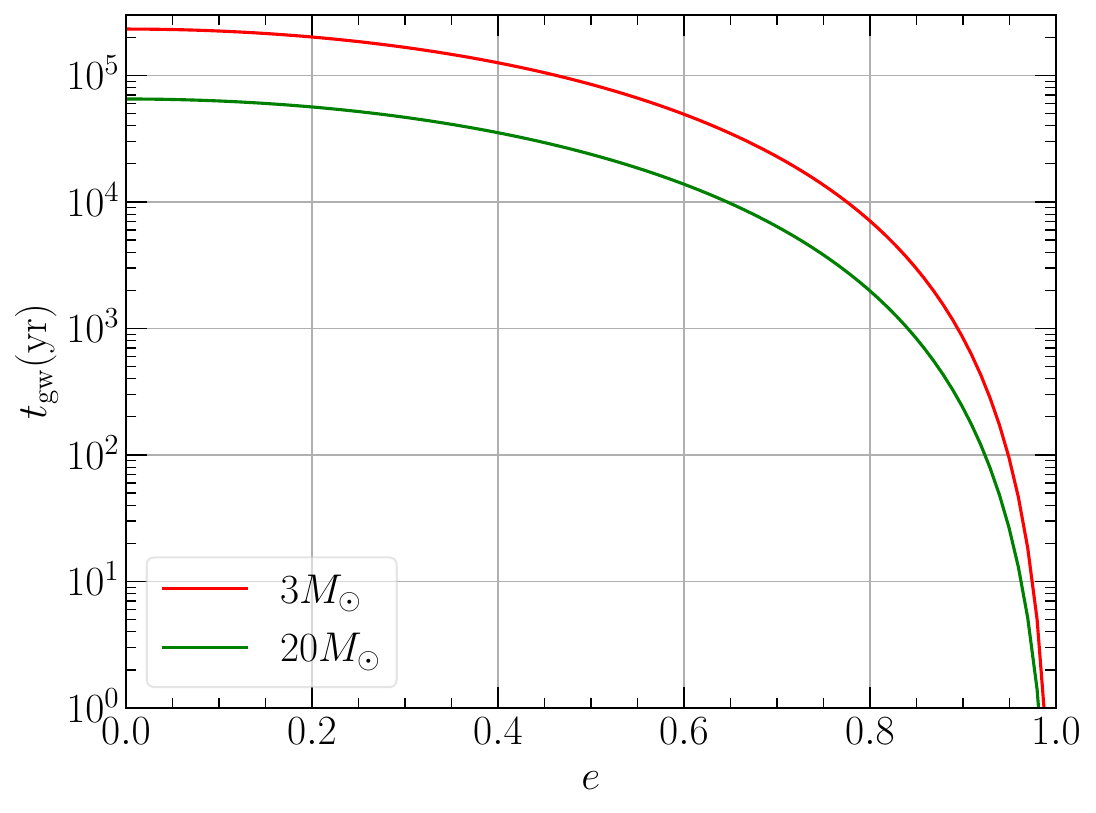}

\caption{The merger time as the function of $e$ for $m_1=3M_\odot$(red) and $20M_\odot$(green).
}
\label{fig:t_e}
\end{figure}

According to Kepler's third law,
\begin{equation}
GM=a^3(2\pi f_{\rm orb})^2,
\end{equation}
we can solve out semimajor axis $a$ from this Equation.
For circular orbit, the merger time $t_{\rm gw}$ (upper panel, see Eq.\eqref{eq:t_m}) and semimajor axis $a$ (lower panel) as the function of $m_1$ is shown in Figure~\ref{fig:ta_m}. Its semimajor axis $a\sim10^{-3}$\,AU.  The merger time $t_{\rm gw}$ decreases with primary mass $m_1$. 

If its orbit is eccentric, the timescale for an eccentric binary with semimajor axis $a$ and eccentricity $e$ decaying to merger by the GW radiation is \citep{1964PhRv..136.1224P,maggiore2008gravitational},
\begin{eqnarray}
t_{\rm gw}(a,e) = 2.32\times10^{5}{\rm yr}\left(\frac{m_1}{3M_\odot}\right)^{-2/3} F(e),
\label{eq:t_gw}
\end{eqnarray}
where $T_{\rm orb}$ is the orbital period of the BBH,
\begin{equation}
F(e)=\frac{48}{19} \frac{1}{\tilde{a}^4(e)} \int^{e}_0 de' \frac{\tilde{a}^4(e')(1-e'^2)^{5/2}}{e'\left(1+\frac{121}{304}e'^2\right)},
\end{equation}
and
\begin{equation}
\tilde{a}(e)=\frac{e^{12/19}}{1-e^2} \left(1+\frac{121}{304} e^2 \right)^{870/2299}.
\end{equation}
the GW decay timescale decreases with increasing mass ratios $m_1$.
For general eccentric orbit, the merger time $t_{\rm gw}$ as the function of $e$ is shown in Figure~\ref{fig:t_e} for $m_1=3M_\odot$(red) and $20M_{\odot}$(green).

\bibliographystyle{yahapj}
\bibliography{refer}
\end{CJK*}
\end{document}